\documentclass[emulateapj,twocolumn]{aastex63}

\usepackage{amsmath}
\usepackage{natbib}
\usepackage{mathtools}
\usepackage{float}
\usepackage{bm}
\restylefloat{table}

\newcommand{\be}{\begin{equation}} 
\newcommand{\ee}{\end{equation}}

\newcommand{\ba}{\begin{eqnarray}} 
\newcommand{\ea}{\end{eqnarray}}

\newcommand{\D}{{\rm d}}

\newcommand{\rin}{r_{\rm in}}
\newcommand{\rout}{r_{\rm out}}

\newcommand{\thetabd}{\theta_{\rm bd}}
\newcommand{\thetasb}{\theta_{\rm sb}}

\newcommand{\wbd}{\omega_{\rm bd}}
\newcommand{\wsb}{\omega_{\rm sb}}
\newcommand{\wbs}{\omega_{\rm bs}}

\newcommand{\tdisk}{t_{\rm disk}}

\newcommand{\fb}{f_{\rm b}}

\newcommand{\Lb}{L_{\rm b}}
\newcommand{\Ld}{L_{\rm d}}

\newcommand{\Md}{M_{\rm d}}
\newcommand{\Mb}{M_{\rm b}}
\newcommand{\ab}{a_{\rm b}}

\newcommand{\Jvec}{{\bm J}}
\newcommand{\Lbvec}{{\bm L}_{\rm b}}
\newcommand{\Ldvec}{{\bm L}_{\rm d}}
\newcommand{\ldvec}{{\bm {\hat l}}_{\rm d}}
\newcommand{\lbvec}{{\bm {\hat l}}_{\rm b}}
\newcommand{\jvec}{\bm {\hat j}}
\newcommand{\svec}{\bm {\hat s}}

\newcommand{\xvec}{\bm {\hat x}}
\newcommand{\yvec}{\bm {\hat y}}
\newcommand{\zvec}{\bm {\hat z}}

\newcommand{\msun}{M_\odot}
\newcommand{\rsun}{R_\odot}

\shorttitle{Obliquity Excitation in Binaries}
\shortauthors{K. R. Anderson \& D. Lai}

\begin{document} 

\title{Excitation of Spin-Orbit Misalignments in Stellar Binaries with Circumbinary Disks: Application to DI Herculis}

\correspondingauthor{Kassandra Anderson}
\email{kassandra@princeton.edu}
\author{Kassandra R. Anderson} 
\affiliation{Department of Astrophysical Sciences, Princeton University, Princeton, NJ 08544, USA}
\author{Dong Lai} 
\affiliation{Cornell Center for Astrophysics and Planetary Science, Department of Astronomy, Cornell University, Ithaca, NY 14853, USA}
\affiliation{Tsung-Dao Lee Institute, Shanghai Jiao Tong University, Shanghai 200240, China}

\begin{abstract}
The large spin-orbit misalignments in the DI Herculis stellar binary system have resolved the decades-long puzzle of the anomalously slow apsidal precession rate, but raise new questions regarding the origin of the obliquities. This paper investigates obliquity evolution in stellar binaries hosting modestly-inclined circumbinary disks. As the disk and binary axes undergo mutual precession, each oblate star experiences a torque from its companion star, so that the spin and orbital axes undergo mutual precession. As the disk loses mass through a combination of winds and accretion, the system may be captured into a high-obliquity Cassini state (a spin-orbit resonance). The final obliquity depends on the details of the disk dispersal. We construct a simple disk model to emulate disk dispersal due to viscous accretion and photoevaporation, and identify the necessary disk properties for producing the observed obliquities in DI Herculis. The disk must be massive (at least $10 \%$ of the binary mass). If accretion onto the binary is suppressed, the observed high stellar obliquities are reproduced with a binary-disk inclination of $\sim 5^\circ - 10^\circ$, but if substantial accretion occurs, the inclination must be larger, $\sim 20^\circ - 30^\circ$. If moderate accretion occurs, initially the disk must lose its mass slowly, but eventually lose its remaining mass abruptly, analogous to the observed two-timescale behavior for disks around T-Tauri stars. The spin feedback on the binary orbit causes the binary-disk inclination to decay as the obliquity evolves, a feature that is absent from the standard Cassini state treatment.
\end{abstract}


\section{Introduction}
\label{sec:intro}
A natural expectation of stellar binary formation is an orbital angular momentum axis aligned with the stellar spin axes, reflecting the rotation axis of the initial proto-stellar cloud.  Indeed, stellar binaries with semi-major axes less than about $40$ AU may exhibit evidence for spin-orbit alignment, whereas wider binaries are more randomly oriented (\citealt{hale1994}; but see \citealt{justesen2020}). In any case, the existence of closely-orbiting (with semi-major axis $< 1$ AU) stellar binaries with significantly misaligned sky-projected spin and orbital axes is of great interest. Such misalignments may suggest an anomaly in the usual star formation process, or post-formation dynamical interactions.  This paper focuses on the latter process.

An extreme example of a binary exhibiting spin-orbit misalignment is DI Herculis, which consists of a pair of B stars (masses $5.15 \msun$ and $4.52 \msun$) in a $10.5$ day orbit \citep{popper1982}. From the Rossiter-McLaughlin effect, \cite{albrecht2009} obtained sky-projected obliquities of $72^\circ$ and $-84^\circ$ for the primary and secondary respectively, and the 3D configuration of the system was further constrained by gravity-darkening \citep{philippov2013}. Prior to the obliquity measurements by \cite{albrecht2009}, DI Herculis had gained considerable attention over several decades, due to an observed apsidal precession rate that appeared to be in contradiction with general relatively \citep{martynov1980, guinan1985,claret1998}. Assuming spin-orbit alignment, the net apsidal precession rate (due to contributions from general relativity, tidal distortion and stellar oblateness) appeared to be too slow compared with theoretical predictions. The large obliquities in the DI Herculis system resolve the longstanding puzzle of the apsidal precession by reversing the direction of the contribution from stellar oblateness, as predicted by \cite{shakura1985}. However, a new puzzle has emerged regarding the origin of the obliquities.

An unseen tertiary companion within several AU could be responsible for generating the observed obliquities \citep[see Fig.~13 of][]{anderson2017a}. In this paper we discuss another possible mechanism for the nearly perpendicular spins of the DI Herculis components, due to the presence of an inclinated circumbinary disk.  The binary-disk interaction introduces nodal precession of the binary orbital axis. Meanwhile, each oblate star experience a torque from its partner star, causing mutual precession of the spin and binary axes.  As the circumbinary disk evolves due to a combination of accretion and winds, a secular spin-orbit resonance may be encountered (equivalently, the system may become captured into a Cassini state; e.g.~\citealt{ward2004}). As the disk mass continues to decrease with time, the system tracks the Cassini state, causing the obliquities to approach $90^\circ$ under some circumstances.

This paper evaluates the prospects for reproducing the observed obliquities in DI Herculis, starting from spin-orbit alignment. We identify the necessary conditions for sustained obliquity growth, accounting for the mutual torques within the spin-binary-disk system, and allowing for mass accretion from the disk onto the binary in a parameterized manner. 

In recent years, the dynamical role of dissipating protostellar disks on orbital architectures has been recognized, and secular resonances similar to the one discussed in this paper have been invoked to explain various observed features of exoplanetary systems \citep{batygin2013, lai2014,zanazzi2018,petrovich2019}. During the completion of this work, a similar mechanism to the one discussed in this paper (for exciting stellar obliquities) was independently proposed by \cite{millholland2019} for exciting planetary obliquities. However, in this paper, we highlight a rich dynamical feature not often appreciated in previous work: We demonstrate the importance of the spin feedback on the binary orbit, in spite of the fact that the spin angular momentum is much smaller than the binary and (initial) disk angular momenta. In contrast, the standard treatment of Cassini states assumes negligible spin. 

This paper is organized as follows: In Section \ref{sec:obliquity} we review the dynamics of mutually inclined spin-binary-disk systems, including ``generalized'' Cassini states, which differ from the classical Cassini-state problem by accounting for the finite spin angular momentum. We then demonstrate how a dispersing disk may generate the observed obliquities in DI Herculis. In Section \ref{sec:accretion} we discuss how mass accretion from the disk onto the binary may under some circumstances suppress obliquity growth. We then identify the the necessary initial disk properties that may reproduce the observed obliquities in the DI Herculis system, using a disk dispersal model motivated by viscous accretion and photoevaporation.  We conclude in Section \ref{sec:conclusion}.

\section{Spin-Orbit Dynamics} \label{sec:obliquity}
\subsection{Setup and Equations}
We consider a binary with component masses $m_0$ and $m_1$, reduced mass $\mu_{\rm b}$, total mass $\Mb$, semi-major axis $\ab$ and angular momentum $\Lb = \mu_{\rm b} \sqrt{G \Mb \ab}$.  An inclined circumbinary disk has mass $\Md$, and inner and outer radii $\rin$ and $\rout$, with $\ab < \rin \ll \rout$.  We adopt a power-law surface density profile, with 
\be
\Sigma (r) = \Sigma_{\rm out} \bigg(\frac{r}{\rout} \bigg)^{-3/2} \simeq \frac{\Md}{4 \pi \rout^2} \bigg(\frac{r}{\rout} \bigg)^{-3/2}.
\label{eq:sigma}
\ee
The angular momentum of the disk, $\Ld$, is given by
\be
\Ld \simeq \frac{\Md}{2} \sqrt{G \Mb \rout}.
\ee
We allow the disk mass to decrease with time according to
\be
\Md (t) = \frac{M_{\rm d,0}}{1 + t/t_{\rm disk}},
\label{eq:Mdisk}
\ee
where $M_{\rm d,0}$ and $t_{\rm disk}$ are free parameters. The total mass-loss rate of the disk is therefore
\be
\dot{M}_{\rm d} = - \frac{M_{\rm d,0}}{t_{\rm disk} (1 + t/t_{\rm disk})^{2}}.
\ee

Due to their mutual inclination, the binary axis $\lbvec = \Lbvec / \Lb$ and disk axis $\ldvec = \Ldvec / \Ld$ undergo mutual precession. The precession rate of $\lbvec$ around $\ldvec$ is $\wbd (\lbvec \cdot \ldvec)$, with
\be
\omega_{\rm b d} =  \frac{3}{20}\frac{\Md}{\Mb} \frac{\ab^3}{\rout^3} \varepsilon^{-5/2} n_{\rm b},
\label{eq:wbd}
\ee
where $n_{\rm b} = \sqrt{G \Mb /\ab^3} = 2 \pi /P_{\rm b}$ is the binary mean motion, and $\varepsilon \equiv \rin / \rout$. Equation (\ref{eq:wbd}) can be derived by calculating the torque on the binary due to the quadrupole-order perturbation from each mass element ${\rm d} M_{\rm d} (r)$ in the disk, integrating over the entire disk, and relating the torque to the change in angular momentum of the binary \citep[see, e.g.][]{owen2017}. The precession rate of $\ldvec$ around $\lbvec$ is $\omega_{\rm db} (\lbvec \cdot \ldvec)$, with $\omega_{\rm db} = (L_{\rm b} / L_{\rm d}) \omega_{\rm bd}$.
Equivalently, both $\lbvec$ and $\ldvec$ precess around their total angular momentum axis $\jvec = \Jvec/J = (\Lbvec + \Ldvec)/J$. The precession of $\lbvec$ around $\jvec$ has the rate
\ba
\omega_{\rm b j} & = & \frac{J}{\Ld} \omega_{\rm b d} (\lbvec \cdot \ldvec) \\ \nonumber
& \simeq & \frac{3}{10} \frac{\mu_{\rm b}}{\Mb} \bigg(\frac{\ab}{\rout} \bigg)^{7/2} \varepsilon^{-5/2} n_{\rm b} (\lbvec \cdot \ldvec) \quad {\rm for} \quad  \Lb \gg \Ld.
\label{eq:wbj}
\ea
In the limit $\Lb \gg \Ld$, $\omega_{\rm bj}$ is independent of the disk mass. Thus, for a fixed disk profile, $\omega_{\rm bj}$ settles to a constant as the disk loses mass. Note that throughout this paper, we treat the disk as a ``rigid plate'', i.e.~it has a negligible warp. This is a reasonable approximation for protostellar disks, because bending waves can efficiently communicate different regions of the disk \citep[e.g.][]{foucart2014,zanazzi2018b}.

Meanwhile, due to the stellar oblateness, the spin axis of $m_0$ ($\svec = {\bf S}/S$) experiences a torque from $m_1$\footnote{Throughout this paper, we consider only the torque on the oblate $m_0$; identical expressions for $m_1$ are obtained by switching the indices ``0'' and ``1''.}, and precesses around $\lbvec$ with the characteristic frequency
\be
\omega_{\rm s b} = \frac{3 k_q}{2 k_\star} \frac{m_1}{m_0} \left(\frac{R_\star}{\ab} \right) ^3 \Omega_\star,
\ee
where $R_\star$ and $\Omega_\star$ are the stellar radius and rotation frequency, and $k_\star \simeq 0.06$ and $k_q \simeq 0.01$ are dimensionless constants describing the interior structure of the star\footnote{$k_\star$ and $k_q$ are defined through the stellar moment of inertia and quadrupole moment: $I_3 = k_\star M_\star R_\star^2$, and $I_3 - I_1 = k_q \hat{\Omega}_\star^2 M_\star R_\star^2$, with $\hat{\Omega}_\star = \Omega_\star (G M_\star / R_\star^3)^{-1/2}$.}. The binary experiences a backreaction torque from $m_0$, causing $\lbvec$ to precess around $\svec$ with much lower frequency $\omega_{\rm bs} = (S/L_{\rm b}) \omega_{\rm sb}$.
We neglect the small coupling between the stellar spin and the disk.

The secular equations of motion for the spin, binary, and disk unit vectors, encapsulating the mutual torques are thus
\ba
\frac{\D \svec}{\D t} & = & \omega_{\rm s b} (\svec \cdot \lbvec) (\svec \times \lbvec), \nonumber \\
\frac{\D \lbvec}{\D t} & = & \omega_{\rm bd} (\lbvec \cdot \ldvec) (\lbvec \times \ldvec) +  \omega_{\rm bs} (\svec \cdot \lbvec) (\lbvec \times \svec), \nonumber \\
\frac{\D \ldvec}{\D t} & = &  \omega_{\rm db} (\lbvec \cdot \ldvec) (\ldvec \times \lbvec)
\label{eq:torques}
\ea

The resulting spin-orbit dynamics depend on the relative precession frequencies and angular momentum ratios \citep{henrard1987,ward2004,boue2014,lai2018}.

\subsection{Generalized Cassini States}
As the disk disperses, the binary-disk precession frequencies change with time, allowing for the possibility of capture into a high obliquity Cassini state. Cassini states are equilibrium states of the vector trio $\svec$, $\lbvec$ and $\ldvec$. We orient $\lbvec$ along the $\zvec$-axis (i.e. $\lbvec = \zvec$) and place the disk axis in the $\yvec$-$\zvec$ plane.  The unit vectors $\ldvec$ and $\svec$ can be written as
\ba
\ldvec & = & \sin \thetabd \yvec + \cos \thetabd \zvec, \\
\svec & = & \sin \thetasb \sin \phi \xvec - \sin \thetasb \cos \phi \yvec + \cos \thetasb \zvec,
\ea
where the obliquity and binary-disk inclination are defined through $\cos \thetasb = \svec \cdot \lbvec$ and $\cos \thetabd = \lbvec \cdot \ldvec$ respectively, and $\phi$ is the phase of $\svec$ around $\lbvec$. In terms of the unit vectors, $\phi$ can be expressed as
\be
\tan \phi = \frac{\svec \cdot (\lbvec \times \ldvec)}{\svec \cdot \ldvec - (\svec \cdot \lbvec)(\lbvec \cdot \ldvec)}.
\ee
The equilibrium states occur when the relative orientations of all three axes are fixed in time \citep[see also][]{boue2006,fabrycky2007cass,correia2015,anderson2018}, and satisfy
\ba
\big[ \svec \cdot (\lbvec \times \ldvec) \big] & = & 0, \nonumber \\
\frac{\D }{\D t} \big[ \svec \cdot (\lbvec \times \ldvec) \big] & = & 0.
\ea
The first condition implies that $\phi = 0, \pi$, i.e., the three axes are coplanar.  The second condition specifies the Cassini state angles, which becomes \citep{anderson2018}
\ba
&& \frac{\omega_{\rm bd}}{\omega_{\rm sb}} \cos \thetabd \bigg[ \cos \thetabd \cos (\thetasb - \thetabd) - \cos \thetasb \bigg] \nonumber \\ 
&& + \frac{S}{\Lb} \cos \thetasb \bigg[ \cos \thetabd - \cos (\thetasb - \thetabd) \cos \thetasb \bigg] \nonumber \\ 
&& - \sin \thetabd \sin \thetasb \bigg[ \cos \thetasb - \frac{\Lb \omega_{\rm bd}}{\Ld \omega_{\rm sb}} \cos \thetabd \bigg] = 0.
\label{eq:cassini}
\ea
After specifying values of the precession frequencies and angular momentum ratios, along with a binary-disk inclination $\thetabd$, the Cassini state obliquity may be obtained by numerically solving equation (\ref{eq:cassini}) for $\theta_{\rm sb}$.

Alternatively, the Cassini states can be thought of as an ordered pair of the obliquity and binary-disk inclination $(\thetasb, \thetabd)$, which depends on the total angular momentum of the system and precession frequencies. Following \cite{correia2016}, we define the angular momentum constant $K_0$:
\ba
K_0 & = & S \Lb \cos \thetasb + S \Ld \cos \theta_{\rm sd} + \Lb \Ld \cos \thetabd \nonumber \\
& = & \frac{1}{2} \big(K^2 - S^2 - \Lb^2 - \Ld^2 \big).
\label{eq:K0}
\ea
where ${\bf K} = {\bf S} + {\bf L}_{\rm b} + {\bf L}_{\rm d}$.
Since,
\be
\cos \theta_{\rm sd} = \cos \thetasb \cos \thetabd - \sin \thetasb \sin \thetabd \cos \phi ,
\ee
equation (\ref{eq:K0}) can be solved as (\citet{correia2016}, equations 78-80)
\be
\cos \thetabd = \frac{Z(\Lb + S \cos \thetasb) - S \sin \thetasb \cos \phi \sqrt{1 - Z^2} }{G},
\label{eq:cosI}
\ee
where 
\ba
Z & = & \frac{K_0 - S \Lb \cos \thetasb}{\Ld G} \\
G & = & \sqrt{(\Lb + S \cos \thetasb)^2 + S^2 \sin^2 \thetasb \cos^2 \phi}.
\ea
Setting $\phi = 0$ or $\pi$, equations (\ref{eq:cassini}) and (\ref{eq:cosI}) specify the ordered pairs of Cassini state obliquity and inclination, given values of $\omega_{\rm sb}, \omega_{\rm bd}, S,\Lb,\Ld,K_0$. These equations must be solved numerically. For the purposes of the numerical root-finding algorithm there is a maximum possible value of $\theta$ $(\theta_{\rm max})$, from requiring that $Z^2 < 1$, given by
\be
\cos \theta_{\rm max} = \frac{K_0 + \Ld^2 - \Ld \sqrt{2 K_0 + S^2 + \Lb^2 + \Ld^2}}{S \Lb}.
\ee

The number of Cassini states depends on the relative precession rates and the angular momentum ratios.  In the classical Cassini state problem (neglecting the feedback of the spin on the orbit), four Cassini states exist, two of which are stable, denoted as the ordered pairs $(\theta_1, I_1)$ and $(\theta_2, I_2)$.  In the limit of strong spin-orbit coupling ($\omega_{\rm bd} / \omega_{\rm sb} \ll 1$), $\theta_1 \to 0$ and $\theta_2 \to \pi/2$.  In the limit of weak spin-orbit coupling ($\omega_{\rm bd} / \omega_{\rm sb} \gg 1$) only one stable state exists ($\theta_2, I_2$).  Cassini state 1 ceases to exist when the ratio $\omega_{\rm bd} / \omega_{\rm sb}$ is of order unity. See, for example, Fig.~3 of \cite{ward2004} or Fig.~3 of \cite{anderson2018}. When the feedback of the spin on the orbit is included, the behavior of the Cassini states is usually qualitatively similar to the idealized case, with the exceptions that for large values of $S/\Lb$, the limiting Cassini state 2 obliquity may not always approach $\pi/2$ in the limit of strong spin-orbit coupling, and additional Cassini states can emerge (see Fig.~6 of \citealt{anderson2018}).

Beginning with a sufficiently massive disk, the spin and binary orbit will initially be weakly coupled, so that the system may be driven into libration around Cassini state 2 ($\theta_2, I_2$).  As the disk mass decreases, the obliquity and binary-disk inclination will track ($\theta_2, I_2$) as the ratio $\omega_{\rm bd} / \omega_{\rm sb}$ decreases.  Following the dispersal of the disk, the obliquity will freeze to a fixed value corresponding to the asymptotic value of $\theta_2$, in the limit of $\omega_{\rm bd} / \omega_{\rm sb} \ll 1$.

\subsection{Fiducial Parameters \& Relevant Timescales}
\label{sec:parameters}

For the remainder of this paper, we will fix the binary properties to values appropriate for a DI Herculis-type system\footnote{DI Herculis has an observed eccentricity of $0.49$, which we neglect in this paper, since such an eccentricity would only introduces slight changes in $\wsb$ and $\wbd$ of order 1. We note that the response of an eccentric binary to a circumbinary disk is a complicated problem, and detailed modelling of the eccentricity evolution due to the binary-disk torque is beyond the scope of this paper (see the discussion in Section \ref{sec:discuss}).}, choosing $m_0 = m_1 = 5 M_\odot$, $R_\star = 2.5 R_\odot$, $P_{\star} = 1.25$ days \citep{philippov2013}, and $\ab = 0.2$ AU. To identify and characterize the key dynamical features, we will first simplify the problem by ignoring the mass accretion onto the binary. We thus begin by holding the stellar masses constant and neglecting possible changes in the orientations of the spin, binary, and disk axes due to accretion. More realistic models are explored in Section \ref{sec:accretion}.

In order for the system to become captured into Cassini state 2, the disk perturbation must initially be sufficiently strong so that the low-obliquity Cassini state 1 does not exist.  As a result, an approximate requirement for obliquity excitation is $\omega_{\rm bd} / \omega_{\rm sb} \gtrsim 1$ initially.  This places constraints on the necessary disk properties, with a minimum disk mass (in units of the binary mass) given by
\ba
&& \frac{M_{\rm d,0}}{\Mb} \gtrsim 0.21 \lambda \frac{m_1}{m_0} \left(\frac{\Mb}{10 \msun} \right)^{-1/2} \left(\frac{\ab}{0.2 {\rm AU}} \right)^{-3/2} \nonumber \\
&& \times \left(\frac{R_\star}{2.5 \rsun} \right)^{3}  \left(\frac{P_\star}{{\rm 1.25 d}} \right)^{-1} \bigg(\frac{\rin}{2 \ab} \bigg)^3 \bigg(\frac{\varepsilon}{10^{-2}} \bigg)^{-1/2},
\label{eq:Mdmin}
\ea
where $\lambda \equiv 6 k_q / k_\star \sim 1$. Equation (\ref{eq:Mdmin}) shows that for DI Herculis-type systems, $\omega_{\rm bd} / \omega_{\rm sb} \gtrsim 1$ is satisfied for an initial disk mass of order $10 \%$ of the binary mass or larger.  For the remainder of Section \ref{sec:obliquity}, we will fix the disk properties to $\rin = 2 \ab$, $\rout = 100 \rin$, and initial mass $M_{\rm d,0} = 0.5 \Mb = 5 \msun$, so that equation (\ref{eq:Mdmin}) is comfortably satisfied. 

\begin{figure}
\centering 
\includegraphics[scale=0.65]{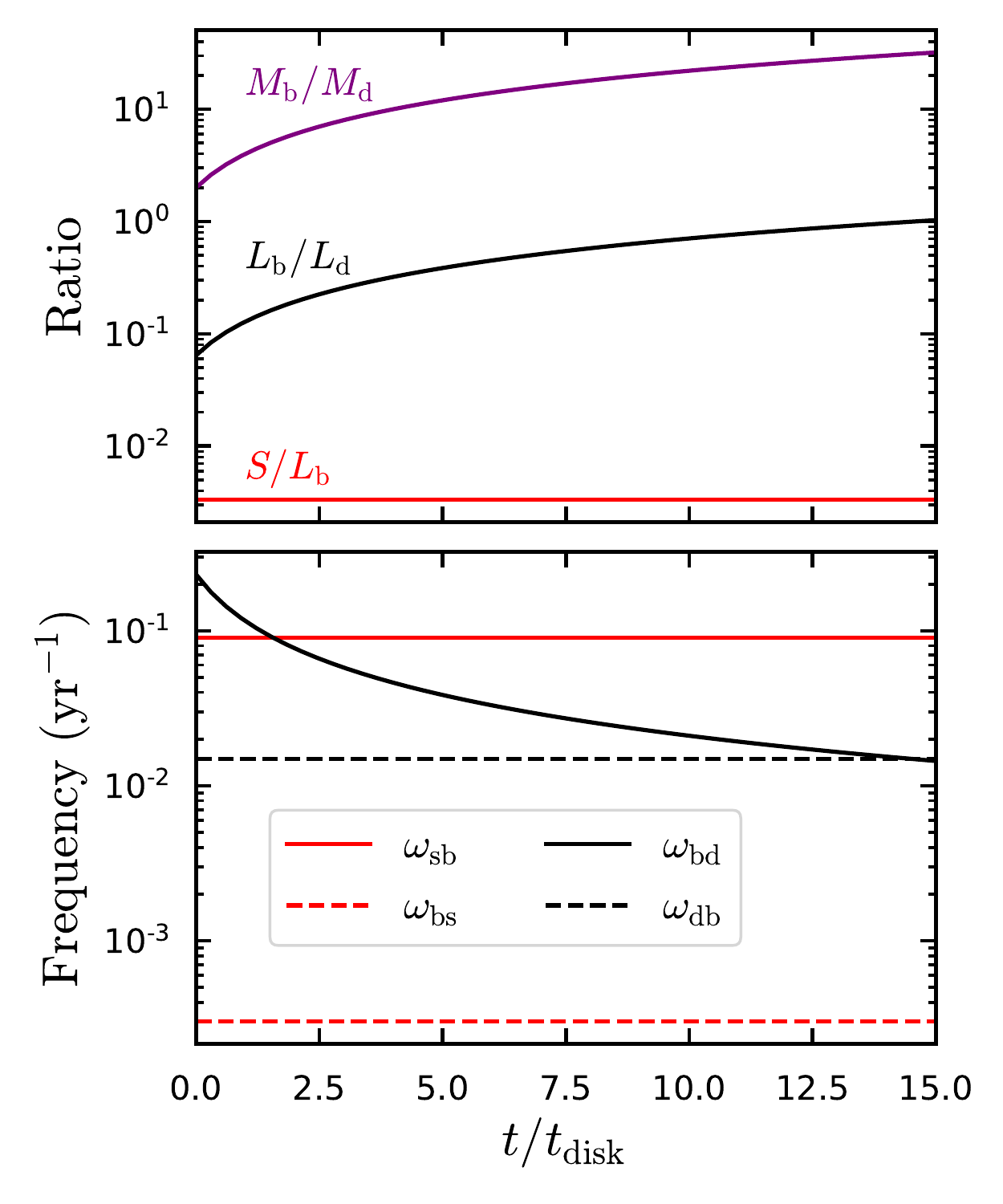}
\caption{Mass and angular momentum ratios (top) and precession frequencies (bottom) as a function of time. The binary parameters have been set to values appropriate for DI Herculis ($m_0 = m_1 = 5 \msun$, $R_\star = 2.5 \rsun$, $\ab = 0.2$ AU), and the disk properties are $\rin = 2 a_{\rm b}$, $\rout = 100 \rin$, and initial mass $M_{\rm d,0} = 0.5 M_{\rm b} = 5 \msun$. These fiducial binary and disk parameters are adopted in Figs.~\ref{fig:example_feedback} and \ref{fig:inc_feedback}.}
\label{fig:timescales}
\end{figure}

In order for the system to be permanently trapped in Cassini state 2, the mass loss must be ``adiabatic,'' with the mass loss timescale $\tau_{\rm d} \equiv M_{\rm d} / |\dot{M}_{\rm d}|$ sufficiently large compared to other timescales in the problem. For the power-law mass loss law (equation \ref{eq:Mdisk}), $\tau_{\rm d}$ increases with time. As a result, if the evolution is adiabatic at the beginning of the disk evolution, it will be satisfied for all times. Inspecting the precession frequencies in the bottom panel of Fig.~\ref{fig:timescales} indicates that $\tdisk \gtrsim 10^{3}$ years will satisfy the adiabatic condition. In this paper we adopt $\tdisk = 10^4$ years to speed up numerical calculations, and note that identical results are obtained for larger $\tdisk$.

For other mass-loss laws, such as an exponential decay law (so that $\tau_{\rm d} = {\rm constant}$), adiabaticity may be satisfied at the beginning of the evolution but not at later times. By requiring that $\tau_{\rm d} \gtrsim 2 \pi / \omega_{\rm bd}$, adiabaticity is maintained as long as the disk mass satisfies
\be
\frac{\Md}{\Mb} \gtrsim 10^{-3} \bigg(\frac{\rin}{2 \ab} \bigg)^{3} \bigg(\frac{\varepsilon}{10^{-2}} \bigg)^{-1/2} \bigg(\frac{P_{\rm b}}{10 {\rm d}} \bigg) \bigg(\frac{\tau_{\rm d}}{10^4 {\rm yr}} \bigg)^{-1}.
\label{eq:adiabatic}
\ee
As a result, for the fiducial parameters, adiabaticity may only be broken when the disk mass is small. 

Fig.~\ref{fig:timescales} illustrates how the precession frequencies, binary-disk mass ratio, and angular momentum ratios evolve with time, assuming our fiducial parameters. Inspecting the top panel, the disk angular momentum is initially greater than the binary angular momentum by a factor of $\sim 10$.  As the disk mass decreases, the ratio $\Lb / \Ld$ increases and reaches unity after a time $15 \tdisk$ has elapsed.  Since we expect the disk homology to change as the disk mass becomes small, in this section we do not consider the disk evolution for times beyond $15 \tdisk$. 

Inspecting the bottom panel of Fig.~\ref{fig:timescales}, the binary nodal precession frequency around the disk $(\wbd)$ is initially greater than the stellar spin axis precession frequency around the binary axis ($\wsb$), so that the spin is relatively weakly coupled to the orbit.  When the disk mass becomes small, the spin becomes strongly coupled to the binary orbit.  Since $\Lb \gg S$, the binary axis precesses with a very low frequency $\wbs = (S / \Lb) \wsb \ll \wsb$. 

\subsection{Examples: Importance of the Spin Feedback on the Binary Orbit}
\label{sec:feedback}
\begin{figure*}
\centering 
\includegraphics[width=0.85\textwidth]{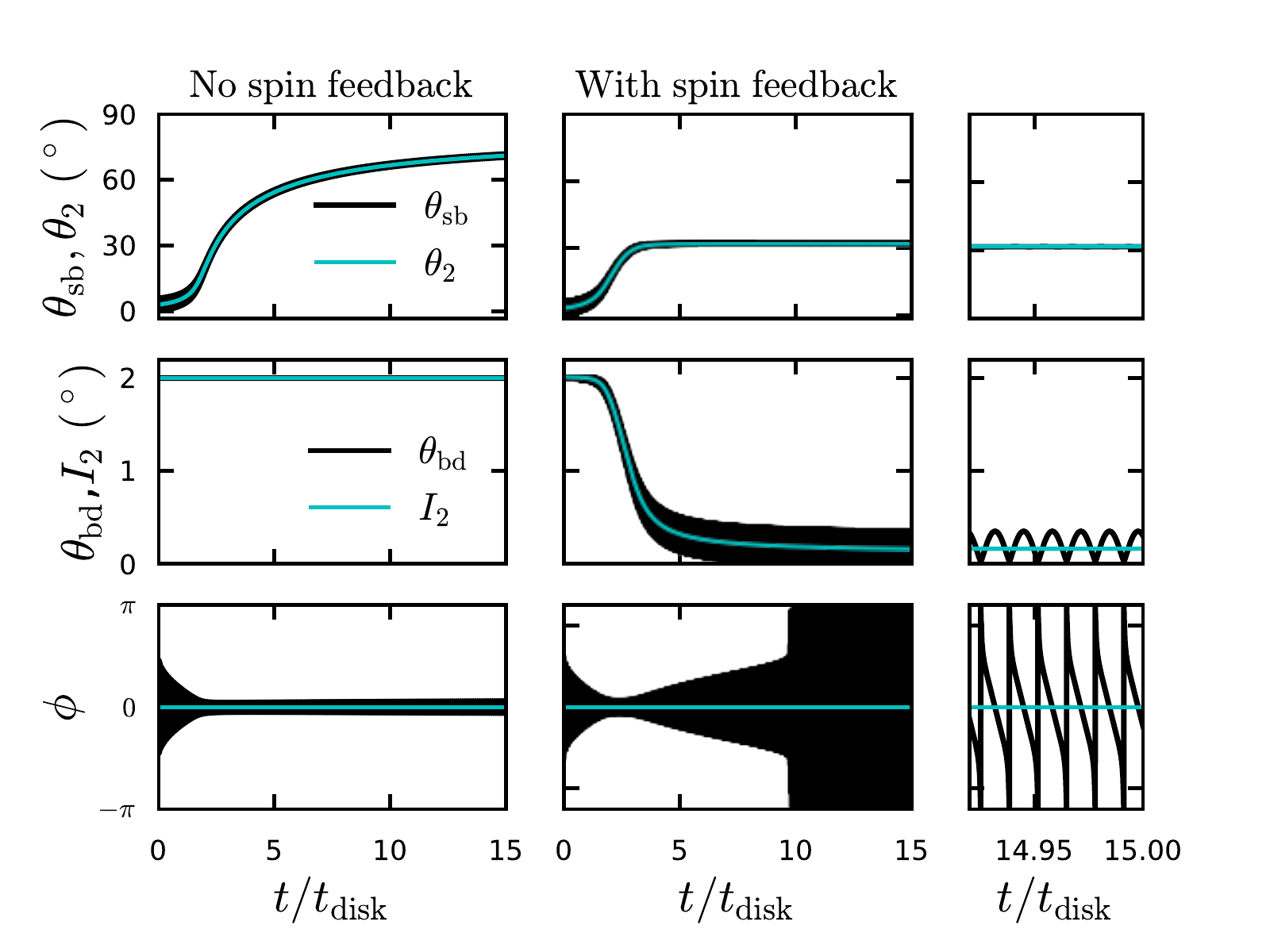}
\caption{Obliquity $(\thetasb)$, binary-disk inclination $(\thetabd)$, and spin axis phase angle $(\phi)$ versus time, adopting the fiducial parameters (as in Fig.~\ref{fig:timescales}).  The results from the numerical integration of equations (\ref{eq:torques}) are shown in black, and the instantaneous pair of Cassini state 2 angles ($\theta_2, I_2$) are calculated analytically from the total angular momentum obtained from the numerical integration (shown in cyan). The left panels show the result without including the spin feedback on the orbit (i.e.~setting $\wbs = 0$ in equation \ref{eq:torques}), and the middle and right panels show the same example with the spin feedback included ($\wbs \neq 0$). We see that although $S / L_{\rm b} \ll 1$ so that $\wbs \ll \wbd$ (see Fig.~\ref{fig:timescales}), including the spin feedback has a dramatic effect, causing $\theta_{\rm bd}$ to damp to nearly zero, and the final obliquity to settle to a value far below $90^\circ$.}
\label{fig:example_feedback}
\end{figure*}

Next, we conduct numerical integrations of the spin-binary-disk system depicted in Fig.~\ref{fig:timescales}, and show the results in Fig.~\ref{fig:example_feedback}. Since $S/\Lb \ll 1$, it is reasonable to begin by neglecting the spin-feedback on the binary orbit (artificially setting $\wbs = 0$ in equation \ref{eq:torques}). Using the total angular momentum obtained from the numerical integration as a function of time, we calculate the instantaneous Cassini state obliquity and inclination $(\theta_2, I_2)$ using equations (\ref{eq:cassini}) and (\ref{eq:cosI}). As expected, the system tracks the Cassini state. Starting with an initial binary-disk inclination $\theta_{\rm bd,0} = 2^\circ$, the system is quickly captured into libration around Cassini state 2, and steadily tracks this equilibrium state to a large obliquity $(\thetasb \simeq 70^\circ)$, as shown in the left panels of Fig.~\ref{fig:example_feedback}. The binary-disk inclination remains constant, due to the fact that the binary experiences no torque from the star, and simply precesses around the disk. 

Notice that in this example, we have only integrated the system up to a time $15 \tdisk$. The system may continue to evolve at later times, but over such timescales the assumption of a fixed disk profile may become progressively problematic. For very low disk masses, holding the disk surface density profile and inner and outer disk radii fixed can cause the obliquity to stall at $\thetasb < 90^\circ$. This occurs because in the limit $\Lb \gg \Ld$, the relevant precession frequency in determining the Cassini states is $\omega_{\rm bj}$, which becomes independent of the disk mass (see equation [\ref{eq:wbj}]). Thus, as $\Md \to 0$, $\omega_{\rm bj} \to {\rm constant}$ in this simple disk model. Clearly such behavior is merely an artifact of holding the disk homology fixed. In Section \ref{sec:photoevap} we explore a more realistic disk model, constructed such that $\omega_{\rm bj} \to 0$ as $\Md \to 0$.

The right-hand panels of Fig.~\ref{fig:example_feedback} show results for the same system parameters, but now including the spin feedback on the binary orbit (so that $\omega_{\rm bs} \neq 0$). In spite of the fact that $S / L_{\rm b} \ll 1$, including the spin feedback causes $\thetabd$ to decay to less than $0.1^\circ$. At such a low inclination, the corresponding Cassini state 2 obliquity is only $\theta_2 \simeq 31^\circ$. Thus, the spin feedback on the binary orbit can drastically affect the final obliquity and cannot be neglected for these parameters.

\begin{figure}
\centering 
\includegraphics[scale=0.6]{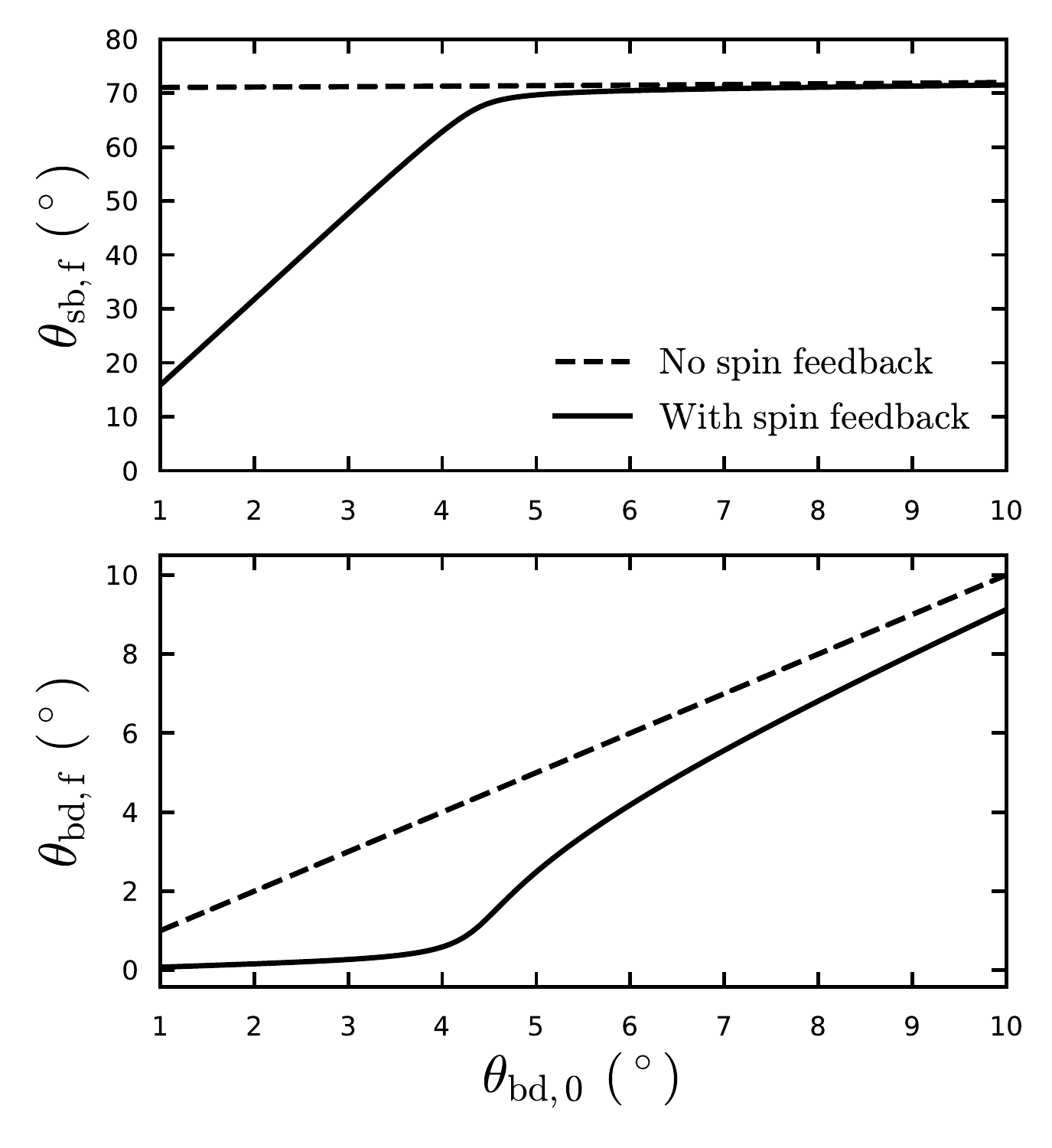}
\caption{Final obliquity (top panel) and binary-disk inclination (bottom panel) at a time $t = 15 \tdisk$. The calculations are the same as in Fig.~\ref{fig:example_feedback}, but for various values of $\theta_{\rm bd,0}$. The solid lines show results including the spin feedback (finite $\wbs$), and the dashed lines show results with the spin feedback neglected (by setting $\wbs = 0$).   For small initial binary-disk inclinations ($\theta_{\rm bd,0} \lesssim 4.5^\circ$), inclusion of the spin feedback can drastically reduce the final inclination, thereby reducing the final obliquity.}
\label{fig:inc_feedback}
\end{figure}

Next, we repeat the calculations shown in Fig.~\ref{fig:example_feedback}, but varying the initial binary-disk inclination in the range $1^\circ - 10^\circ$. As before, we compare the results with and without spin feedback included. Fig.~\ref{fig:inc_feedback} shows the final obliquity (top panel) and inclination (bottom panel) at a time $15 \tdisk$. For initial binary-disk inclinations $\theta_{\rm bd,0} \lesssim 4.5^\circ$, the spin-feedback dramatically reduces the binary-disk inclination, leading to a final obliquity far less than $90^\circ$.

In summary, we have demonstrated that at small inclinations, the backreaction torque from the oblate star on the orbit plays an essential role in the spin-binary-disk dynamics, leading to damping of the mutual inclination, and a final obliquity that can be far less than $90^\circ$. Somewhat counter-intuitively (since $S / \Lb \ll 1$), the spin-feedback on the binary orbit often cannot be neglected, and the classical Cassini state treatment is invalid.

\section{Effects of Accretion onto the Binary} \label{sec:accretion}
\subsection{Accretion Torques}
We now allow for the possibility of accretion onto the binary. The total mass accreted onto the binary is specified as
\be
\dot{M}_{\rm acc,b} = \fb |\dot{M}_{\rm d}|,
\ee
with $\fb$ a free parameter between $0-1$.  Since we consider an equal-mass binary, we assume each star accretes at the rate $\dot{m}_0 = \dot{m}_1 = \fb |\dot{M}_{\rm d}|/2$. In this section we will hold the stellar masses fixed even for non-zero $\fb$ to make the problem cleaner, but allow the stellar masses to increase in Section \ref{sec:photoevap}.

The mass accreted from the misaligned disk onto the binary introduces torques, which act to align the disk and binary axes, as well as the spin and binary axes. We write the accretion torques as
\ba
\bigg( \frac{\D {\bm {L}}_{\rm b}}{\D t} \bigg)_{\rm acc} & = & N_{\rm bd} \ldvec, \nonumber \\
\bigg( \frac{\D {\bm {S}}}{\D t} \bigg)_{\rm acc} & = & N_{\rm sb} \lbvec.
\label{eq:torguevec}
\ea
with 
\ba
N_{\rm bd} & \simeq & \dot{M}_{\rm acc,b} \sqrt{G \Mb \ab}, \nonumber \\
N_{\rm sb} & \simeq & \dot{M}_{\rm acc,s} \sqrt{G m_0 R_\star}.
\label{eq:aligntorques}
\ea
Note that the precise values for the torques in equation (\ref{eq:aligntorques}) may be smaller by a factor of $\sim 2$ \cite[see][]{miranda2017,munoz2019,munoz2020}. We ignore such a correction in this paper, as this correction can be absorbed into the accretion efficiency factor $\fb$. Equation (\ref{eq:torguevec}) implies that the alignment torques on $\lbvec$ and $\svec$ are
\ba
\bigg( \frac{\D \lbvec}{\D t} \bigg)_{\rm acc} & = & \frac{N_{\rm bd}}{\Lb} \bigg(\ldvec - \cos \theta_{\rm bd} \lbvec \bigg), \nonumber \\
\bigg( \frac{\D \svec}{\D t} \bigg)_{\rm acc} & = & \frac{N_{\rm sb}}{S} \bigg(\lbvec - \cos \theta_{\rm sb} \svec \bigg).
\label{eq:accretion}
\ea
We add equations (\ref{eq:accretion}) to the expressions for $\D \lbvec / \D t$ and $\D \svec / \D t$ in equations (\ref{eq:torques}). For simplicity, we ignore the evolution of $\Lb$ and $S$. This is because other torques may change the angular momentum magnitudes (e.g.~accretion onto the star may lead to magnetic stellar wind torques that balance the accretion torque). In any case, we do not expect the evolution of $S$ and $\Lb$ to significantly influence the evolution of $\svec$ and $\lbvec$. In all remaining calculations, we include the feedback of the spin on the binary orbit.

\begin{figure*}
\centering 
\includegraphics[width=\textwidth]{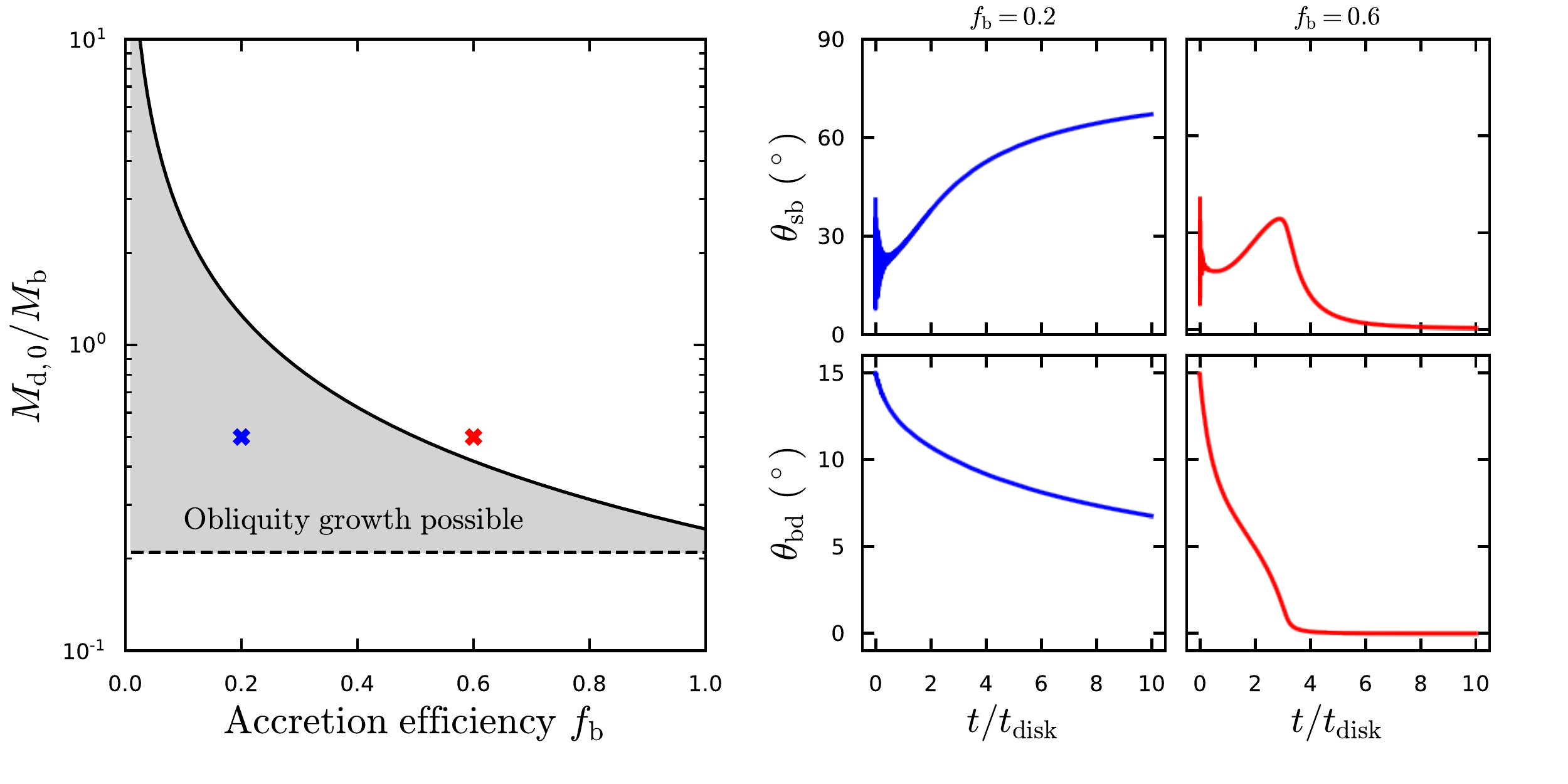}
\caption{{\it Left:} Estimated parameter space allowing sustained obliquity excitation. The solid curve indicates the maximum initial disk mass needed to prevent binary-disk realignment (equation \ref{eq:Mdmax}), while the dashed line indicates the minimum initial disk mass needed for capture into Cassini state 2 (equation \ref{eq:Mdmin}). The colored crosses indicate the adopted parameters in the time-evolution examples depicted in the right-hand panels. {\it Right panels}: Examples of the obliquity evolution (top) and binary-disk inclination (bottom) for suppressed accretion ($f_{\rm b} = 0.2$), and moderate accretion ($f_{ \rm b} = 0.6$).}
\label{fig:example_acc}
\end{figure*}

Sustained obliquity excitation requires that the binary and disk maintain a sufficiently high inclination through time.  In order to prevent immediate alignment of $\lbvec$ and $\ldvec$, the alignment timescale must be greater than the disk mass loss timescale: $L_{\rm b} / N_{\rm bd} \gtrsim M_{\rm d} / |\dot{M}_{\rm d}|$.  This implies a maximum initial disk mass in order to prevent alignment:
\be
M_{\rm d,0} \lesssim \frac{\mu_{\rm b}}{f_{\rm b}}.
\label{eq:Mdmax}
\ee
There also exists a minimum initial disk mass necessary for capture into Cassini state 2 (equation [\ref{eq:Mdmin}]).  Together, these conditions specify a range of initial disk masses allowing obliquity excitation due to capture into Cassini state 2, but preventing immediate star-binary-disk alignment due to accretion.  This range of disk masses is shown in the left panel of Fig.~\ref{fig:example_acc}, assuming the canonical binary and disk parameters.  At large $f_{\rm b}$, the available parameter space for obliquity growth narrows.

The right panels in Fig.~\ref{fig:example_acc} show the obliquity and inclination evolution for two different accretion efficiencies: Suppressed accretion ($f_{\rm b} = 0.2$), and moderate accretion ($f_{\rm b} = 0.6$). For the suppressed accretion example, the system retains a binary-disk inclination above a few degrees, leading to a $\sim 75^\circ$ final obliquity.  In contrast, in the moderate accretion example ($f_{\rm b} = 0.6$), the accretion torques act to quickly align the entire spin-binary-disk system.

The left panel of Fig.~\ref{fig:example_acc} roughly delineates the parameter space for obliquity excitation to be possible, but does not yield information on the actual degree of obliquity growth, or whether the obliquity can be maintained with time. Thus, the boundaries in Fig.~\ref{fig:example_acc} are a necessary, but not sufficient condition for sustained obliquity growth. Indeed, if the $\fb = 0.2$ example shown in Fig.~\ref{fig:example_acc} is integrated three times longer, the binary-disk inclination is eventually damped to less than $1^\circ$, causing the obliquity to settle to $\sim 55^\circ$. However, the simplified model for disk dispersal adopted (equation \ref{eq:Mdisk}) is not expected to hold for arbitrary times. The final obliquity thus depends on the details of the disk dispersal, and we explore this issue in Section \ref{sec:photoevap}. 

\subsection{Dependence on the Disk Dispersal: Viscous Evolution \& Photoevaporation}
\label{sec:photoevap}
In the previous examples we have adopted a simple model for the disk dispersal, by specifying the mass-loss rate through equation (\ref{eq:Mdisk}) and holding the disk homology fixed. We now explore a more realistic disk dispersal model.

Disk dispersal occurs due to a combination of viscous evolution and disk winds/photoevaporation. Significant uncertainties remain regarding the relative importance of these dispersal agents. Photoevaporation can occur due to the high-energy radiation emitted by the stellar binary members, as well as that from the external environment. In the case of internal photoevaporation (due to the host binary), the evolution may proceed qualitatively similarly to the photevaporation of circumstellar disks around isolated stars \citep{alexander2012}. High-energy photons are incident upon the inner disk, heating the gas so that the sound speed exceeds the escape velocity at a distance of $\sim $ several AU from the host star. A photoevaporative wind thus develops in concert with inward viscous accretion, causing an inner hole to develop. Eventually the hole completely cuts off mass supply from the outer disk, so that the inner disk drains onto the host star. At this point, the weakly-bound mass in the outer disk is exposed to the full amount of radiation from the star and is rapidly destroyed from the inside-out, over a timescale much shorter than the disk lifetime. The well-known ``two-timescale'' behavior for disk dispersal around observed T-Tauri stars \cite[e.g.][]{andrews2005,alexander2014} is consistent with the predictions of internal photoevaporation \citep{clarke2001}. Similar dispersal processes may occur for disks around massive binaries such as DI Herculis. 

If the external FUV field is sufficiently strong, a photoevaporative flow is initiated in the outer disk \citep[e.g.][]{adams2004}. The external radiation acts to truncate the size of the disk, causing the overall viscous timescale to be short, so that the disk quickly disperses due to a combination of mass accretion onto the host star and winds in the outer disk. The disk may thus first be destroyed from the outside-in before the radiation from the host is able to destroy the disk \citep{anderson2013}. We will not explore this scenario further in this paper, but simply point out the possibility for this qualitatively distinct dispersal process.

We adopt a toy-model to emulate the dispersal of the disk due to photevaporation from the stellar hosts, by requiring that the disk mass initially decrease slowly with time according to a power-law (as in equation [\ref{eq:Mdisk}]), before rapidly dispersing on an exponential timescale. The disk mass thus takes the form
\be
\Md (t) = 
\begin{dcases}
\frac{M_{\rm d,0}}{1 + t / \tdisk}, \quad & t \leq t_{\rm crit} \\
\frac{M_{\rm d,0}}{1 + t_{\rm crit} / t_{\rm disk}} \exp{\bigg[-\frac{(t - t_{\rm crit})}{\tdisk} \bigg]}, & t > t_{\rm crit}. \\
 \end{dcases}
\label{eq:Mdisk_evap}
\ee
We choose $t_{\rm crit} = 10 \tdisk$. Equation (\ref{eq:Mdisk_evap}) mimics the two-timescale behavior for disk dispersal, by allowing the disk to evolve slowly when $t < t_{\rm crit}$, with a sudden dispersal when $t > t_{\rm crit}$.

To capture the effect of viscous spreading, we allow the disk radius to increase with time according to
\be
\rout (t) = r_{\rm out,0} \bigg(1 + c \frac{t}{\tdisk} \bigg).
\ee
We choose $r_{\rm out,0} = 100 \rin$ and $c = 1$, setting $\rin = 2 \ab$ and fixing the surface density profile to that of equation (\ref{eq:sigma}) as before. By allowing the outer radius to increase with time, the precession frequency $\omega_{\rm bj} \to 0$ as $\Md \to 0$. We therefore avoid the artificial behavior discussed in Section \ref{sec:feedback}. We hold the inner disk radius constant. Although this model is simplistic, it captures the key features of disk evolution due to a combination of viscosity and photoevaporation from the host star. 

Finally, here we allow the binary mass to increase with time due to the mass accretion from the disk according to
\be
\Mb (t) = M_{\rm b,0} + \fb \big[\Md(t) - M_{\rm d,0} \big],
\ee
For a given value of $\fb$, we specify the initial binary mass $M_{\rm b,0}$ such that $\Mb = 10 \msun$ (the total mass of the DI Herculis members) at a time $t = 10 \tdisk$.

To explore the parameter space and ascertain the prospects for obliquity growth, we conduct numerical integrations of a large number of star-binary-disk systems, using the updated disk model and integrating up to a time $15 \tdisk$, at which point the disk mass is negligibly small (see equation \ref{eq:Mdisk_evap}). We explore two different initial binary-disk inclinations: $\theta_{\rm bd,0} = 10^\circ$ and $30^\circ$. For each $\theta_{\rm bd,0}$, we generate 1000 systems, with the accretion efficiency $\fb$ sampled uniformly in $[0,1]$, and the initial disk-binary mass ratio $M_{\rm d,0} / M_{\rm b,0}$ uniformly sampled from $[0.1,1]$. The largest mass ratios are probably unrealistic but included for the purposes of illustration.

\begin{figure*}
\centering 
\includegraphics[width=0.8\textwidth]{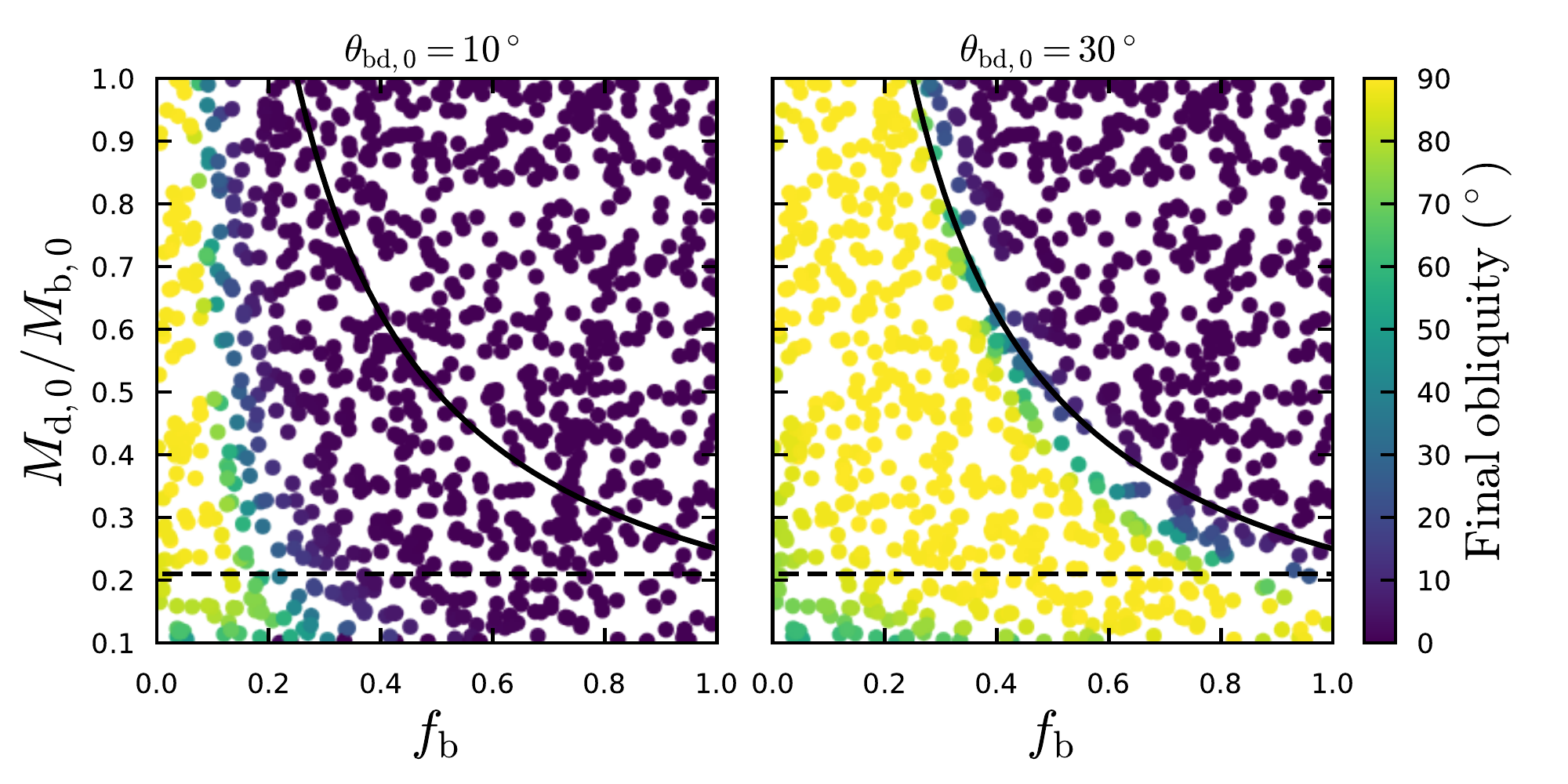}
\caption{Results of a large number of numerical integrations of star-binary-disk systems, exploring a range of accretion efficiencies $\fb$ and initial disk-binary mass ratios $M_{\rm d,0} / M_{\rm b,0}$. The color indicates the obliquity following the disk dispersal (see Section \ref{sec:photoevap}). The solid and dashed lines indicate the estimated parameter space for sustained obliquity growth, as shown in Fig.~\ref{fig:example_acc}. For the $10^\circ$ initial binary-disk inclination, accretion onto the binary must be highly suppressed in order for the obliquity to reach $\sim 90^\circ$. For the $30^\circ$ inclination, $90^\circ$ obliquities may be achieved even when the accretion efficiency $\fb$ is substantial.}
\label{fig:obliquity_fb}
\end{figure*}

Figure \ref{fig:obliquity_fb} shows the final obliquities after the disk dispersal. In the left panel, with $\theta_{\rm bd,0} = 10^
\circ$, obliquity growth is prevented unless accretion onto the binary is highly suppressed ($\fb \lesssim 0.1$). In the right panel, with $\theta_{\rm bd,0} = 30^\circ$, large obliquities can be generated, even for moderate accretion ($\fb \gtrsim 0.5$). The solid and dashed lines indicate the estimated parameter space for sustained obliquity growth, as shown in Fig.~\ref{fig:example_acc}. When $\theta_{\rm bd,0} = 30^\circ$, these estimates well-predict the systems with large final obliquities.

Throughout this paper we have adopted binary properties similar to the observed properties of the DI Herculis system.  However, since the obliquity growth must occur early, the initial binary properties may have differed.  In particular, the initial stellar radii may have been larger or smaller, depending on whether the binary has accreted a significant amount of mass from the disk. Furthermore, the binary semi-major axis may have been larger or smaller, due to the possibilities for the binary to gain or lose angular momentum from the disk (see the discussion in Section \ref{sec:conclusion}, as well as the discussion following equation \ref{eq:accretion}). Given the uncertainties involved, we do not attempt to model either the stellar radius evolution or the possible evolution of the binary semi-major axis.

\begin{figure}
\centering 
\includegraphics[scale=0.65]{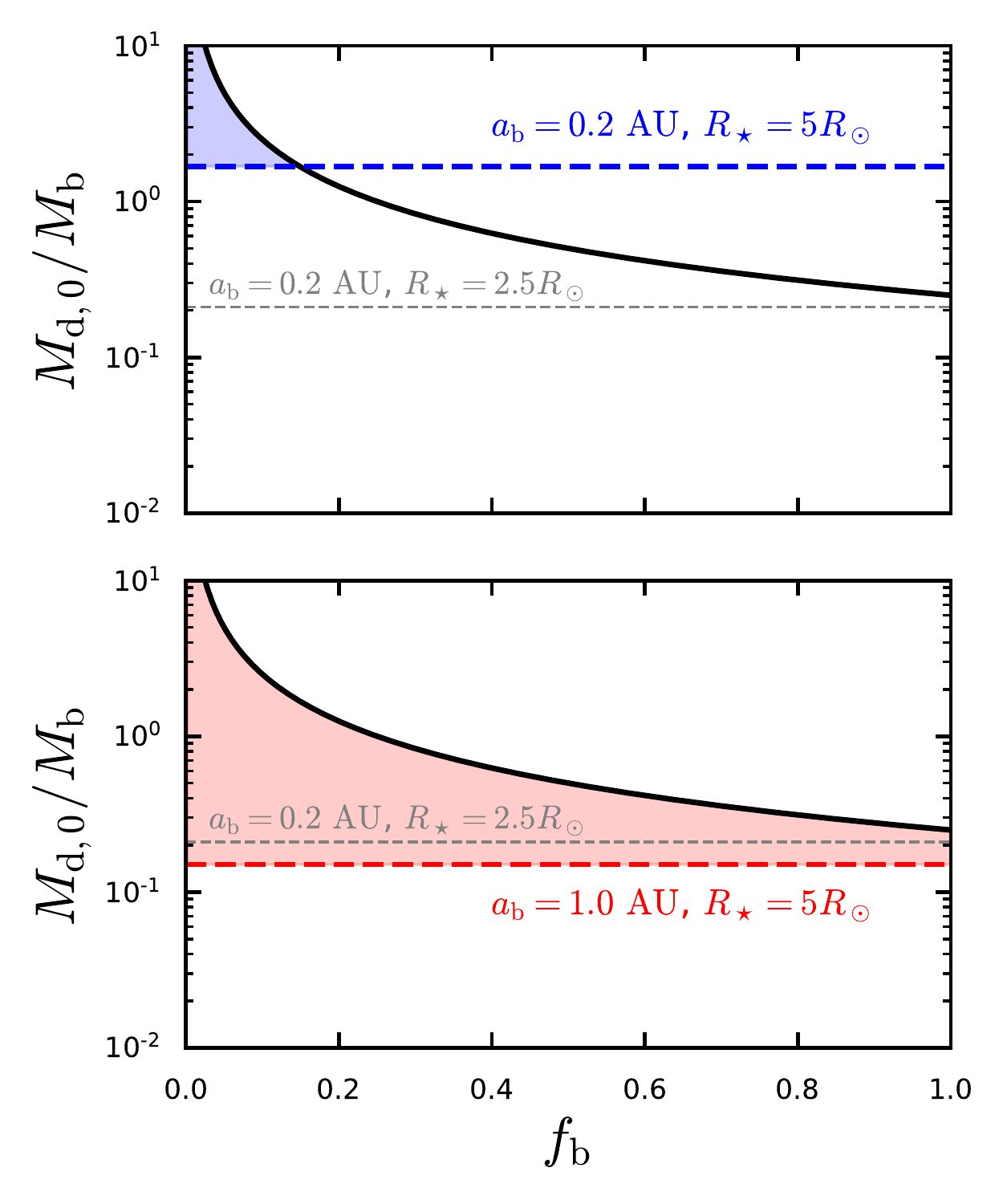}
\caption{Similar to the left panel of Fig.~\ref{fig:example_acc}, illustrating how the stellar radius and binary semi-major axis affect the available parameter space for sustained obliquity excitation.  In both panels, the black curve shows the maximum estimated disk mass to excite the obliquity without realigning the binary and disk (equation [\ref{eq:Mdmax}]), and the dashed lines show the minimum disk mass needed to be captured into Cassini state 2 (equation [\ref{eq:Mdmin}]). The grey dashed lines assume the fiducial binary parameters (as in Fig.~\ref{fig:example_acc}). The red and blue dashed lines show how the minimum disk mass changes with the stellar radius and binary semi-major axis, as labeled. Enlarging the initial stellar radius (top panel) narrows the available parameter space, but enlarging the binary semi-major along with the radius axis widens the parameter space (bottom panel).}
\label{fig:paramspace_panels}
\end{figure}

Figure \ref{fig:paramspace_panels} shows how the parameter space for sustained obliquity excitation widens or narrows with stellar radius and binary semi-major axis. Due to the increase in the stellar quadrupole moment, enlarging the stellar radius by a factor of $2$ significantly narrows the available parameter space for obliquity excitation, and shifts the available parameter space to unrealistically large disk masses. However, enlarging the binary semi-major axis widens the available parameter space. We thus conclude that an enlarged stellar radius may still allow for resonant obliquity growth, provided that the initial binary separation was also larger.

\section{Conclusion}
\label{sec:conclusion}
\subsection{Summary of Results}
Motivated by the large spin-orbit misalignments observed for both members of the DI Herculis stellar binary system, this paper has considered obliquity excitation in stellar binaries due to the presence of an inclined circumbinary disk. While the disk induces nodal precession of the binary orbit, each oblate star experiences a torque from its companion star, causing mutual precession between the spin and binary axes. As the disk mass decreases due to a combination of accretion onto the binary and disk winds, the system may become locked into a particular spin-orbit resonant configuration (Cassini state 2), which may cause the obliquity to be driven towards $90^\circ$. In this paper we identify the disk properties needed to reproduce the nearly polar obliquities observed in the DI Herculis system. We allow for the possibility of alignment torques due to accretion from the disk onto the binary; under some circumstances these torques can suppress obliquity growth. Our main findings are as follows:

\begin{itemize}
    \item In spite of the fact that the stellar spin angular momentum is much smaller than the binary angular momentum, the feedback of each oblate star on the binary orbit can have a dramatic effect on the binary-disk inclination. This feature makes the spin dynamics quite different from the usual Cassini state problem. For initial binary-disk inclinations less than $\sim 5^\circ$, the spin feedback causes the inclination to damp to nearly $0^\circ$, and the obliquity to settle to a value far less than $90^\circ$ (see Figs.~\ref{fig:example_feedback} and \ref{fig:inc_feedback}). 
    
    \item In order to produce the observed obliquities in the DI Herculis system, the initial disk mass must be $\sim 10 \%$ of the initial binary mass or larger. If accretion onto the binary is suppressed (so that the disk mass is lost primarily through photoevaporation/winds), the initial binary-disk inclination must be at least $5^\circ$.
    
    \item Allowing for accretion from the disk onto the binary in a parameterized manner, we find that accretion torques can sometimes suppress obliquity growth by aligning the binary and disk. For a specified accretion efficiency, we estimate the range of necessary binary-disk mass ratios allowing sustained obliquity growth (see Fig.~\ref{fig:example_acc}). If moderate accretion occurs, the initial binary-disk inclination must be sufficiently large ($\sim 20^\circ - 30^\circ$).

\end{itemize}

\subsection{Discussion}
\label{sec:discuss}

Since the model proposed in this paper contains several simplifying assumptions, this paper is meant to serve as a proof-of-concept. Below we discuss several issues that deserve further study.

This paper has neglected changes in the magnitude of the binary angular momentum due to the torque and accretion from the disk. While 1D models of circumbinary disks assume the binary loses angular momentum due to the disk torque (so that the orbital separation shrinks, e.g. \citealt{armitage2002}), recent numerical simulations indicate that the binary gains angular momentum (and hence probably grows) over a wide range of parameters \citep{miranda2017,munoz2019,munoz2020}, although see \cite{tiede2020}. This feature also arises in a 3D numerical study by \cite{moody2019}. Given the uncertainties, here we considered the role of the accretion torques only on the direction of the binary axis. A modest increase in $\ab$ would not change our results, but if the binary does grow significantly, obliquity growth may be suppressed in some cases, due to difficulties in encountering the spin-orbit commensurability responsible for capture into a high-obliquity Cassini state. If the binary shrinks, capture into the high-obliquity Cassini state may be achieved with lower initial disk masses.

Similarly, this paper has neglected the eccentricity evolution of the binary, in spite of the observed $0.5$ eccentricity of DI Herculis. For this eccentricity, the stellar spin and binary nodal precession frequencies would experience only order unity changes. More importantly, highly inclined disks around eccentric binaries may be driven towards a polar orbit \citep{martin2017,zanazzi2018b}. Evaluating equation (8) of \cite{zanazzi2018b} at a binary eccentricity of $0.5$, the critical inclination above which the disk will be driven towards a polar configuration is $I_{\rm crit} \simeq 38^\circ$. Although beyond the scope of this present paper, future treatments of spin-binary-disk systems could account for the simultaneous eccentricity and spin-orbit evolution, allowing for the possibility for the disk to be driven towards polar alignment in highly inclined systems.

Due to the relatively short timescales under consideration, this paper has neglected tidal effects. If tides are sufficiently strong, the obliquity is susceptible to re-alignment following disk dispersal. The observed eccentricity in the DI Herculis system indicates that tides have not had the chance to circularize the system, but the timescale for the obliquity to evolve under tides may be much shorter than the circularization timescale. Using the standard weak-friction theory of tides, and requiring the timescale for obliquity alignment (see, e.g., equation 2 of \citealt{lai2012}) to be longer than the system age ($\sim 5$Myr; \citealt{claret1998}), the modified tidal quality factor $Q'$ must satisfy $Q' \gtrsim 5 \times 10^5$.

The mechanism discussed in this paper may be relevant in other stellar binary systems. Indeed, obliquity excitation due to an inclined disk may occur more naturally in stellar binaries with different masses or architectures. The large masses and rapid rotations of the DI Herculis members introduce large quadrupole moments, so that the spin-axis precession period is only tens of years (see Fig.~\ref{fig:timescales}). Since obliquity excitation requires that the binary nodal precession due to the disk initially be faster than the spin-axis precession, the DI Herculis system requires a massive initial disk. For lower mass stellar binaries, or binaries with wider separations, obliquity excitation may be achieved with smaller disk masses (see equation \ref{eq:Mdmin}). Additional Rossiter-McLaughlin observations of stellar binaries with different masses and orbital separations is therefore of great interest.

The spin-binary-disk dynamics discussed in this paper may have consequences for circumbinary planets. As we have shown, the binary-disk inclination can decrease substantially as the disk evolves due to both the feedback of the spin on the orbit and accretion torques. As a result, planets forming within such aligning disks would have low inclinations relative to the binary orbit, even if an initial binary-disk misalignmnent existed. A similar point was previously made by \cite{correia2016}, who demonstrated that inclined circumbinary planets can be driven towards a coplanar orbit as the stellar spins and binary orbit evolve under tides. Many of the circumbinary planets detected thus far are coplanar with their host binary orbit, although this may arise due to selection biases. 

Since the initial binary-disk inclination tends to be somewhat erased, direct evidence for the obliquity excitation mechanism discussed in this paper may be difficult to obtain. Remnants of the initial gaseous disk may be present in the form of a debris disk, but a present-day coplanar debris disk does not rule out an initially inclined disk. However, indirect evidence for disk excitation may be available in equal-mass binaries. Since stars of similar masses are expected to undergo similar spin-orbit dynamics, a possible signature of obliquity excitation due to a disk could be correlated spin-orbit misalignments in equal-mass binaries, as observed in DI Herculis.

Although this paper has focused on stellar binaries, similar spin dynamics can also operate in star-giant-planet systems. Intriguingly, some hot Jupiters are observed to be in nearly polar orbits relative to their host stars \citep[e.g.][]{albrecht2012,addison2013}, especially for massive hosts. Since disk masses are observed to scale with stellar masses, perhaps super-linearly \citep{pascucci2016}, this finding may indicate that obliquity excitation due to a massive circumstellar disk may be at work in planetary systems\footnote{\cite{petrovich2020} recently proposed a similar mechanism for producing close-in Neptunes in polar orbits.}. However, the large quadrupole moment of the host star during the disk-hosting stage may prevent a polar obliquity from being achieved, by modifying the Cassini state 2 obliquity to a value far below $90^\circ$ (see Fig.~6 of \citealt{anderson2018}). These issues merit further investigation.

\acknowledgments
We thank Ryan Miranda and Josh Winn for useful discussions, and the anonymous referee for comments and feedback. During the completion of this work, KRA was supported by a NASA Earth and Space Science Graduate Fellowship in Planetary Science at Cornell University, and a Lyman Spitzer, Jr. Postdoctoral Fellowship at Princeton University.

\bibliography{refs}

\begin{thebibliography}{}
\expandafter\ifx\csname natexlab\endcsname\relax\def\natexlab#1{#1}\fi
\providecommand{\url}[1]{\href{#1}{#1}}
\providecommand{\dodoi}[1]{doi:~\href{http://doi.org/#1}{\nolinkurl{#1}}}
\providecommand{\doeprint}[1]{\href{http://ascl.net/#1}{\nolinkurl{http://ascl.net/#1}}}
\providecommand{\doarXiv}[1]{\href{https://arxiv.org/abs/#1}{\nolinkurl{https://arxiv.org/abs/#1}}}

\bibitem[{{Adams} {et~al.}(2004){Adams}, {Hollenbach}, {Laughlin}, \&
  {Gorti}}]{adams2004}
{Adams}, F.~C., {Hollenbach}, D., {Laughlin}, G., \& {Gorti}, U. 2004, \apj,
  611, 360, \dodoi{10.1086/421989}

\bibitem[{{Addison} {et~al.}(2013){Addison}, {Tinney}, {Wright}, {Bayliss},
  {Zhou}, {Hartman}, {Bakos}, \& {Schmidt}}]{addison2013}
{Addison}, B.~C., {Tinney}, C.~G., {Wright}, D.~J., {et~al.} 2013, \apjl, 774,
  L9, \dodoi{10.1088/2041-8205/774/1/L9}

\bibitem[{{Albrecht} {et~al.}(2009){Albrecht}, {Reffert}, {Snellen}, \&
  {Winn}}]{albrecht2009}
{Albrecht}, S., {Reffert}, S., {Snellen}, I. A.~G., \& {Winn}, J.~N. 2009,
  \nat, 461, 373, \dodoi{10.1038/nature08408}

\bibitem[{{Albrecht} {et~al.}(2012){Albrecht}, {Winn}, {Butler}, {Crane},
  {Shectman}, {Thompson}, {Hirano}, \& {Wittenmyer}}]{albrecht2012}
{Albrecht}, S., {Winn}, J.~N., {Butler}, R.~P., {et~al.} 2012, \apj, 744, 189,
  \dodoi{10.1088/0004-637X/744/2/189}

\bibitem[{{Alexander}(2012)}]{alexander2012}
{Alexander}, R. 2012, \apjl, 757, L29, \dodoi{10.1088/2041-8205/757/2/L29}

\bibitem[{{Alexander} {et~al.}(2014){Alexander}, {Pascucci}, {Andrews},
  {Armitage}, \& {Cieza}}]{alexander2014}
{Alexander}, R., {Pascucci}, I., {Andrews}, S., {Armitage}, P., \& {Cieza}, L.
  2014, in Protostars and Planets VI, ed. H.~{Beuther}, R.~S. {Klessen}, C.~P.
  {Dullemond}, \& T.~{Henning}, 475,
  \dodoi{10.2458/azu_uapress_9780816531240-ch021}

\bibitem[{{Anderson} {et~al.}(2013){Anderson}, {Adams}, \&
  {Calvet}}]{anderson2013}
{Anderson}, K.~R., {Adams}, F.~C., \& {Calvet}, N. 2013, \apj, 774, 9,
  \dodoi{10.1088/0004-637X/774/1/9}

\bibitem[{{Anderson} \& {Lai}(2018)}]{anderson2018}
{Anderson}, K.~R., \& {Lai}, D. 2018, \mnras, 480, 1402,
  \dodoi{10.1093/mnras/sty1937}

\bibitem[{{Anderson} {et~al.}(2017){Anderson}, {Lai}, \&
  {Storch}}]{anderson2017a}
{Anderson}, K.~R., {Lai}, D., \& {Storch}, N.~I. 2017, \mnras, 467, 3066,
  \dodoi{10.1093/mnras/stx293}

\bibitem[{{Andrews} \& {Williams}(2005)}]{andrews2005}
{Andrews}, S.~M., \& {Williams}, J.~P. 2005, \apj, 631, 1134,
  \dodoi{10.1086/432712}

\bibitem[{{Armitage} \& {Natarajan}(2002)}]{armitage2002}
{Armitage}, P.~J., \& {Natarajan}, P. 2002, \apjl, 567, L9,
  \dodoi{10.1086/339770}

\bibitem[{{Batygin} \& {Adams}(2013)}]{batygin2013}
{Batygin}, K., \& {Adams}, F.~C. 2013, \apj, 778, 169,
  \dodoi{10.1088/0004-637X/778/2/169}

\bibitem[{{Bou{\'e}} \& {Fabrycky}(2014)}]{boue2014}
{Bou{\'e}}, G., \& {Fabrycky}, D.~C. 2014, \apj, 789, 111,
  \dodoi{10.1088/0004-637X/789/2/111}

\bibitem[{{Bou{\'e}} \& {Laskar}(2006)}]{boue2006}
{Bou{\'e}}, G., \& {Laskar}, J. 2006, \icarus, 185, 312,
  \dodoi{10.1016/j.icarus.2006.07.019}

\bibitem[{{Claret}(1998)}]{claret1998}
{Claret}, A. 1998, \aap, 330, 533

\bibitem[{{Clarke} {et~al.}(2001){Clarke}, {Gendrin}, \&
  {Sotomayor}}]{clarke2001}
{Clarke}, C.~J., {Gendrin}, A., \& {Sotomayor}, M. 2001, \mnras, 328, 485,
  \dodoi{10.1046/j.1365-8711.2001.04891.x}

\bibitem[{{Correia}(2015)}]{correia2015}
{Correia}, A. C.~M. 2015, \aap, 582, A69, \dodoi{10.1051/0004-6361/201525939}

\bibitem[{{Correia} {et~al.}(2016){Correia}, {Bou{\'e}}, \&
  {Laskar}}]{correia2016}
{Correia}, A. C.~M., {Bou{\'e}}, G., \& {Laskar}, J. 2016, Celestial Mechanics
  and Dynamical Astronomy, 126, 189, \dodoi{10.1007/s10569-016-9709-9}

\bibitem[{{Fabrycky} {et~al.}(2007){Fabrycky}, {Johnson}, \&
  {Goodman}}]{fabrycky2007cass}
{Fabrycky}, D.~C., {Johnson}, E.~T., \& {Goodman}, J. 2007, \apj, 665, 754,
  \dodoi{10.1086/519075}

\bibitem[{{Foucart} \& {Lai}(2014)}]{foucart2014}
{Foucart}, F., \& {Lai}, D. 2014, \mnras, 445, 1731,
  \dodoi{10.1093/mnras/stu1869}

\bibitem[{{Guinan} \& {Maloney}(1985)}]{guinan1985}
{Guinan}, E.~F., \& {Maloney}, F.~P. 1985, \aj, 90, 1519,
  \dodoi{10.1086/113865}

\bibitem[{{Hale}(1994)}]{hale1994}
{Hale}, A. 1994, \aj, 107, 306, \dodoi{10.1086/116855}

\bibitem[{{Henrard} \& {Murigande}(1987)}]{henrard1987}
{Henrard}, J., \& {Murigande}, C. 1987, Celestial Mechanics, 40, 345,
  \dodoi{10.1007/BF01235852}

\bibitem[{{Justesen} \& {Albrecht}(2020)}]{justesen2020}
{Justesen}, A.~B., \& {Albrecht}, S. 2020, arXiv e-prints, arXiv:2008.12068.
\newblock \doarXiv{2008.12068}

\bibitem[{{Lai}(2012)}]{lai2012}
{Lai}, D. 2012, \mnras, 423, 486, \dodoi{10.1111/j.1365-2966.2012.20893.x}

\bibitem[{{Lai}(2014)}]{lai2014}
---. 2014, \mnras, 440, 3532, \dodoi{10.1093/mnras/stu485}

\bibitem[{{Lai} {et~al.}(2018){Lai}, {Anderson}, \& {Pu}}]{lai2018}
{Lai}, D., {Anderson}, K.~R., \& {Pu}, B. 2018, \mnras, 475, 5231,
  \dodoi{10.1093/mnras/sty133}

\bibitem[{{Martin} \& {Lubow}(2017)}]{martin2017}
{Martin}, R.~G., \& {Lubow}, S.~H. 2017, \apjl, 835, L28,
  \dodoi{10.3847/2041-8213/835/2/L28}

\bibitem[{{Martynov} \& {Khaliullin}(1980)}]{martynov1980}
{Martynov}, D.~I., \& {Khaliullin}, K.~F. 1980, \apss, 71, 147,
  \dodoi{10.1007/BF00646915}

\bibitem[{{Millholland} \& {Batygin}(2019)}]{millholland2019}
{Millholland}, S., \& {Batygin}, K. 2019, \apj, 876, 119,
  \dodoi{10.3847/1538-4357/ab19be}

\bibitem[{{Miranda} {et~al.}(2017){Miranda}, {Mu{\~n}oz}, \&
  {Lai}}]{miranda2017}
{Miranda}, R., {Mu{\~n}oz}, D.~J., \& {Lai}, D. 2017, \mnras, 466, 1170,
  \dodoi{10.1093/mnras/stw3189}

\bibitem[{{Moody} {et~al.}(2019){Moody}, {Shi}, \& {Stone}}]{moody2019}
{Moody}, M. S.~L., {Shi}, J.-M., \& {Stone}, J.~M. 2019, \apj, 875, 66,
  \dodoi{10.3847/1538-4357/ab09ee}

\bibitem[{{Mu{\~n}oz} {et~al.}(2020){Mu{\~n}oz}, {Lai}, {Kratter}, \&
  {Miranda}}]{munoz2020}
{Mu{\~n}oz}, D.~J., {Lai}, D., {Kratter}, K., \& {Miranda}, R. 2020, \apj, 889,
  114, \dodoi{10.3847/1538-4357/ab5d33}

\bibitem[{{Mu{\~n}oz} {et~al.}(2019){Mu{\~n}oz}, {Miranda}, \&
  {Lai}}]{munoz2019}
{Mu{\~n}oz}, D.~J., {Miranda}, R., \& {Lai}, D. 2019, \apj, 871, 84,
  \dodoi{10.3847/1538-4357/aaf867}

\bibitem[{{Owen} \& {Lai}(2017)}]{owen2017}
{Owen}, J.~E., \& {Lai}, D. 2017, \mnras, 469, 2834,
  \dodoi{10.1093/mnras/stx1033}

\bibitem[{{Pascucci} {et~al.}(2016){Pascucci}, {Testi}, {Herczeg}, {Long},
  {Manara}, {Hendler}, {Mulders}, {Krijt}, {Ciesla}, {Henning}, {Mohanty},
  {Drabek-Maunder}, {Apai}, {Sz{\H{u}}cs}, {Sacco}, \&
  {Olofsson}}]{pascucci2016}
{Pascucci}, I., {Testi}, L., {Herczeg}, G.~J., {et~al.} 2016, \apj, 831, 125,
  \dodoi{10.3847/0004-637X/831/2/125}

\bibitem[{{Petrovich} {et~al.}(2020){Petrovich}, {Mu{\~n}oz}, {Kratter}, \&
  {Malhotra}}]{petrovich2020}
{Petrovich}, C., {Mu{\~n}oz}, D.~J., {Kratter}, K.~M., \& {Malhotra}, R. 2020,
  \apjl, 902, L5, \dodoi{10.3847/2041-8213/abb952}

\bibitem[{{Petrovich} {et~al.}(2019){Petrovich}, {Wu}, \&
  {Ali-Dib}}]{petrovich2019}
{Petrovich}, C., {Wu}, Y., \& {Ali-Dib}, M. 2019, \aj, 157, 5,
  \dodoi{10.3847/1538-3881/aaeed9}

\bibitem[{{Philippov} \& {Rafikov}(2013)}]{philippov2013}
{Philippov}, A.~A., \& {Rafikov}, R.~R. 2013, \apj, 768, 112,
  \dodoi{10.1088/0004-637X/768/2/112}

\bibitem[{{Popper}(1982)}]{popper1982}
{Popper}, D.~M. 1982, \apj, 254, 203, \dodoi{10.1086/159724}

\bibitem[{{Shakura}(1985)}]{shakura1985}
{Shakura}, N.~I. 1985, Soviet Astronomy Letters, 11, 224

\bibitem[{{Tiede} {et~al.}(2020){Tiede}, {Zrake}, {MacFadyen}, \&
  {Haiman}}]{tiede2020}
{Tiede}, C., {Zrake}, J., {MacFadyen}, A., \& {Haiman}, Z. 2020, arXiv
  e-prints, arXiv:2005.09555.
\newblock \doarXiv{2005.09555}

\bibitem[{{Ward} \& {Hamilton}(2004)}]{ward2004}
{Ward}, W.~R., \& {Hamilton}, D.~P. 2004, \aj, 128, 2501,
  \dodoi{10.1086/424533}

\bibitem[{{Zanazzi} \& {Lai}(2018{\natexlab{a}})}]{zanazzi2018}
{Zanazzi}, J.~J., \& {Lai}, D. 2018{\natexlab{a}}, \mnras, 478, 835,
  \dodoi{10.1093/mnras/sty1075}

\bibitem[{{Zanazzi} \& {Lai}(2018{\natexlab{b}})}]{zanazzi2018b}
---. 2018{\natexlab{b}}, \mnras, 473, 603, \dodoi{10.1093/mnras/stx2375}

\end{thebibliography}

\end{document}